\documentclass[conference]{IEEEtran}
\makeatletter
\def\ps@headings{%
\def\@oddhead{\mbox{}\scriptsize\rightmark \hfil \thepage}%
\def\@evenhead{\scriptsize\thepage \hfil \leftmark\mbox{}}%
\def\@oddfoot{}%
\def\@evenfoot{}}
\makeatother
\usepackage{graphicx}
\usepackage{float}
\usepackage{epstopdf}
\usepackage{array}

\usepackage{caption}
\usepackage{subcaption}
\usepackage{amsmath}
\usepackage{amsfonts}
\usepackage{algpseudocode}
\usepackage{algorithmicx}
\usepackage{algorithm}
\begin{document}

\title{On Balanced $k$-coverage in Visual Sensor Networks}
\author{
\IEEEauthorblockN{Md. Muntakim Sadik\textsuperscript{1}, Sakib Md. Bin Malek\textsuperscript{2}, Ashikur Rahman\textsuperscript{3}}
\IEEEauthorblockA{\textsuperscript{1,2,3}Department of Computer Science and Engineering,\\
Bangladesh University of Engineering and Technology, Dhaka.\\
Email: \textsuperscript{1}0905003.mms@ugrad.cse.buet.ac.bd, \textsuperscript{2}0905039.smbm@ugrad.cse.buet.ac.bd, \textsuperscript{3}ashikur@cse.buet.ac.bd}
}
\maketitle

\begin{abstract} 
Given a set of directional visual sensors, the $k$-coverage problem  
determines the orientation of minimal directional sensors so that each target is 
covered at least $k$ times. As the problem 
is NP-complete, a number of heuristics have been devised to 
tackle the issue. However, the existing heuristics provide imbalance 
coverage of the targets--some targets are covered $k$ times while others are 
left totally uncovered or singly covered. 
The coverage imbalance is more serious in under-provisioned 
networks where there do not exist enough sensors to cover all 
the targets $k$ times. Therefore, we address the problem of covering 
each target at least $k$ times in a balanced way using minimum number 
of sensors. We study the existing Integer Linear Programming (ILP) 
formulation for single coverage and extend the idea for $k$-coverage. 
However, the extension does not balance the coverage of the targets.
We further propose Integer Quadratic Programming (IQP)  and
Integer Non-Linear Programming (INLP) formulations 
that are capable of addressing the coverage balancing.
As the proposed formulations are computationally expensive, we devise a 
faster Centralized Greedy $k$-Coverage Algorithm (CGkCA) to approximate 
the formulations. Finally, through rigorous simulation experiments
we show the efficacy of the proposed formulations 
and the CGkCA.
\end{abstract}

\section{Introduction}\label{intro}

A visual sensor network (VSN), also known as a Smart Camera Network (SCN), 
consists of a set of targets to be 
monitored by a set of \emph{smart} (visual) sensors capable of self-controlling 
their orientations and ranges.
Such VSNs have drawn considerable attention of researchers due to their enormous applicability 
in real-world scenarios like surveillance system, environment monitoring, smart 
traffic controlling system etc., to name a few.

The primary goal of VSNs is to monitor as many targets as possible \cite{fan2010coverage}, 
\cite{huang2005coverage}. However, if the sensor covering a target malfunctions, runs out of 
power, or if the line of sight is blocked by a perpetrator, a previously covered 
target may suddenly become uncovered. The simplest solution to this problem
is to incorporate \emph{fault tolerance} besides \emph{coverage}, 
i.e., to cover the target by more than one sensor. This joint \emph{fault tolerant coverage}
problem is well-known as ``\emph{$k$-coverage}'' problem in the literature.

Formally, our work tackles the $k$-coverage problem, where each target is to be 
covered by at least $k$ sensors ($k \geq 1$). The efficiency of the solution 
to the problem 
depends on the extent of camera usage as fewer number of active sensors 
implies lower energy consumption and longer network life time. Thus, 
in $k$-coverage problem one needs to minimize
camera-usage besides covering each target in a fault-tolerant way. 
\begin{figure}
        \centering
		\begin{subfigure}[!]{.22\textwidth}
			\centering        
     		\includegraphics[width=\textwidth]{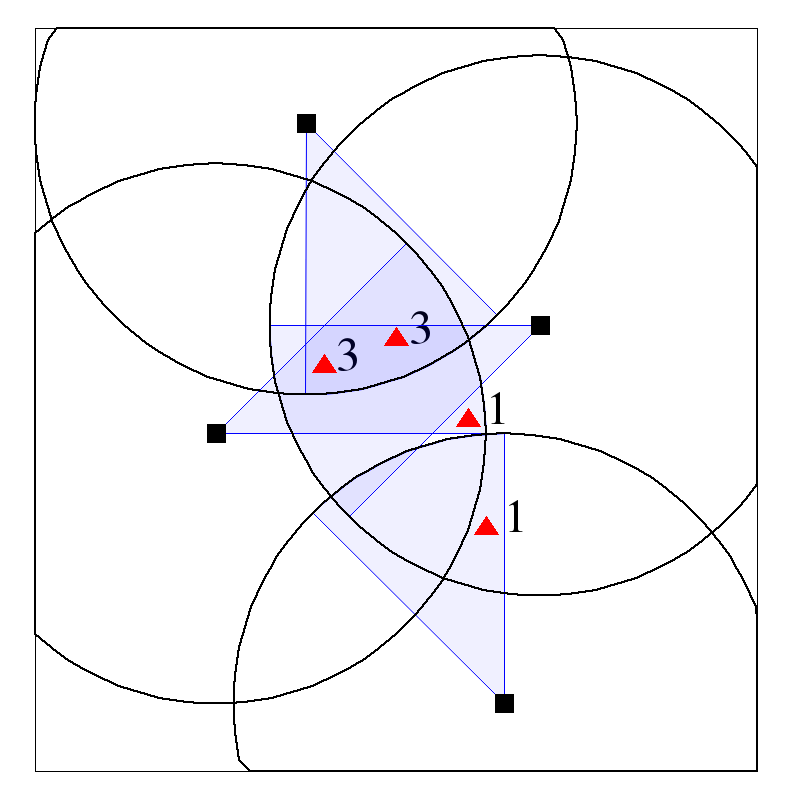}
        	\caption{Imbalanced coverage}
	        \label{fig:sub3}
	    \end{subfigure}    
		\hfill
        \begin{subfigure}[!]{.22\textwidth}
			\centering        
     		\includegraphics[width=\textwidth]{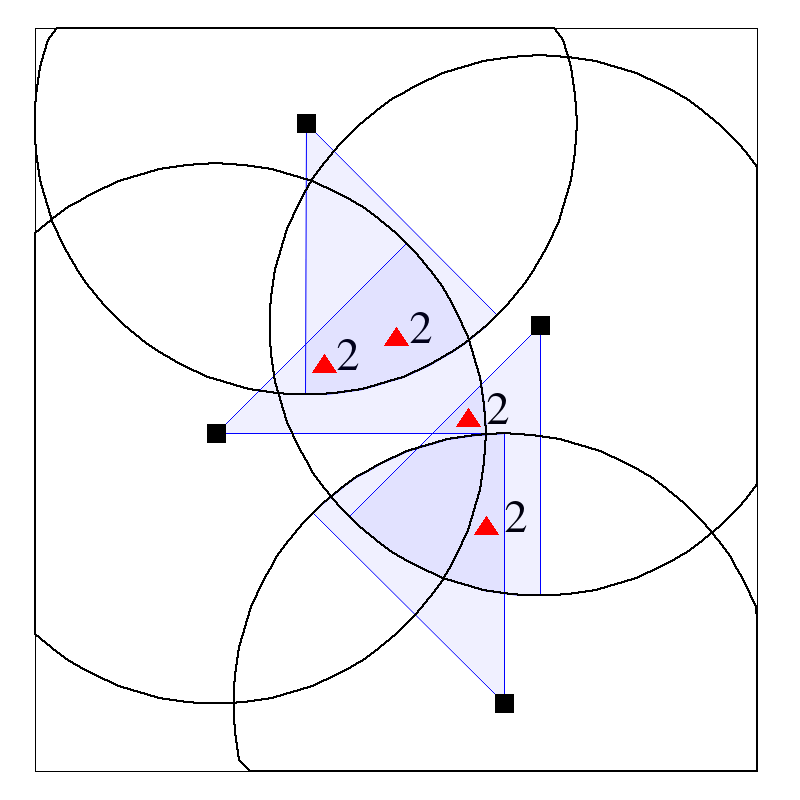}
        	\caption{Balanced coverage}
	        \label{fig:sub4}
	    \end{subfigure}
	    \caption{Coverage imbalance in under-provisioned systems}\label{fig:balanced}
\vskip -0.2in
\end{figure}

We envision two kinds of visual sensor networks:–--(i) under-provisioned networks and (ii) 
over-provisioned networks. We call a VSN is under-provisioned if the number of 
sensors is insufficient to cover all the targets at least $k$ times and over-provisioned 
otherwise. Consider Fig. \ref{fig:sub3} for an example of an under-provisioned network. In this figure, 
there are 4 cameras (rectangular ones) and 4 targets (triangular ones) 
and the objective is to cover each target at least $3$ times (i.e., $k=3$). Here, every 
camera possesses a specific number of non-overlapping pans, of which only one can be 
selected in a particular deployment and each pan is defined by a field of view (FoV) 
angle,  $\theta = \frac{\pi}{4}$. 
Thus, a camera can pick any one of eight $(\frac{2\pi}{\theta} = \frac{2\pi}{\frac{\pi}{4}} = 8)$ 
disjoint pans/orientations. Note that, no orientation of 4 cameras can produce a $3$-coverage 
for this scenario, hence the network is consequently called ``under-provisioned''.

When it is impossible to provide a complete $k$-coverage for an under-provisioned network, 
we need to provide a more \emph{fault resilient} solution instead. For example, let's 
re-consider the under-provisioned network in Fig. \ref{fig:sub3} and Fig. \ref{fig:sub4}. 
Although both the figures show the same deployment of cameras and targets but the orientation 
of the cameras are different. In Fig. \ref{fig:sub3} the targets are covered by 
$3$, $3$, $1$, and $1$ sensors respectively, while in Fig. \ref{fig:sub4}, 
each of the targets are covered by 2 sensors. However, the coverage 
in Fig. \ref{fig:sub4} is more \emph{fault resilient} than the coverage in 
Fig. \ref{fig:sub3} because in Fig. \ref{fig:sub3} two targets are covered 
once, while in Fig. \ref{fig:sub4} all the targets are covered twice. 
We say the coverage in Fig. \ref{fig:sub4} is more \emph{balanced} than 
coverage in Fig. \ref{fig:sub3}.

In this paper, we focus on under-provisioned systems and provide solutions for \emph{coverage balancing} 
using minimum number of sensors. The problem is eventually an instance of the classical  
set \emph{multi-cover} problem whose optimization version is known to be NP-hard~\cite{ai2006coverage}. 
In order to define \emph{coverage balancing}, we borrow the concept of \emph{fairness} in resource allocation 
systems proposed by Jain et al. \cite{jain1998quantitative} and modify it for our purpose 
(the modification is described in a later section). 
The \emph{Fairness Index}, $\mathcal{F}\mathcal{I}$, is defined as follows \cite{jain1998quantitative}.
Suppose we have $1,2,3, .. ,m$ components in a system and $x_i$ is the resource allocated to the 
$i^{th}$ component. The fairness index of such system will be:
\begin{equation}
\mathcal{F}\mathcal{I}  =  \dfrac{ (\sum x_i)^2 }{m \sum x_i^2}
\end{equation}

In VSN, the sensors are the resources and the targets are the components of the system. 
Therefore, we can use the fairness index to judge by how much a system is balanced. Here, the 
number of times each target is covered is considered as the number of resources allocated to that target. 
Thus, the fairness index value of Fig. \ref{fig:sub3} is:
$$\mathcal{F}\mathcal{I} = \dfrac{(3 + 3 + 1 + 1)^2}{4 \times (3^2 + 3^2 + 1^2 + 1^2)} = 0.8$$
and, for Fig. \ref{fig:sub4} is: 
$$\mathcal{F}\mathcal{I} = \dfrac{(2 + 2 + 2 + 2)^2}{4 \times (2^2 + 2^2 + 2^2 + 2^2)} = 1.0$$
Consequently, the camera orientations in Fig. \ref{fig:sub4} has more coverage balancing than the 
camera orientations in Fig. \ref{fig:sub3}. 

The major contributions of this paper are as follows:

(i) We introduce a novel \emph{balanced k-coverage problem} for visual sensor networks (VSNs). 

(ii) We study the existing exact Integer Linear Programming (ILP) formulation 
for single coverage and extend the idea for $k$-coverage. Then we show that 
this natural extension provides imbalance coverage of the targets in a sense 
that some targets are covered $k$-times while some targets are left totally 
uncovered or singly covered. The imbalance is more serious in \emph{under-provisioned} 
networks.

(iii) We propose a novel Integer Quadratic Programming (IQP) 
formulation that improves fairness by balancing coverage while 
trying to achieve $k$-coverage. We further improve the fairness 
by providing another novel Integer Non-Linear Programming (INLP) formulation. 

(iv) The proposed ILP, IQP, and INLP formulations  are 
computationally expensive. Therefore, we formulate a novel computationally 
faster Centralized Greedy $k$-Coverage Algorithm (CG$k$CA) to approximate 
our formulations. Finally, we measure the relative performance of our  
formulations and the CG$k$CA algorithm in terms of coverage balancing.

The road map of the paper is as follows. This section introduces the coverage 
balancing problem and the motivation behind this work. 
Section~\ref{background} introduces the description and parameters of a 
Visual Sensor Network and formally defines the problem that we solve in this paper. 
Section~\ref{background} also discusses the shortcomings of Fairness Index in capturing 
coverage balance and introduces a new metric Balancing Index. 
Section~\ref{ILP} shows how to formulate Integer Linear Programming (ILP) for the 
$k$-coverage problem and discusses ILP's incapability in coverage balancing.
Section~\ref{balanced_k} modifies ILP  
to formulate Integer Quadratic Programming (IQP) and Integer Non-Linear Programming (INLP)
that incorporate coverage balancing besides $k$-coverage. 
Section~\ref{heuristics} discusses the Centralized Greedy $k$-Coverage 
Algorithm (CG$k$CA). Section~\ref{results} presents the simulation results 
and analyzes the results. Section~\ref{relate} provides
a brief literature review on the subject matter and finally Section~\ref{conclusion} 
concludes the paper. Throughout the paper we use the terms ``camera'' and ``sensor'' 
interchangeably.

\section{Preliminaries}
\label{background}
In this section we formally introduce the
VSN with relevant parameters and provide a formal description of the 
problem.

\subsection{Visual Sensor Network Description and Parameters}\label{para}
The sensing region of a camera can be characterized by its Field of View (FoV) 
which is defined as the extent of the observable/sensing region
that can be captured at any given direction.
Some cameras come with fixed-FoV and for some, FoVs are adjustable.
The smart cameras used  in current VSNs are known as \emph{Pan-Tilt-Zoom} (PTZ) cameras
where FoV can be self-adjusted in three dimensions: (i) horizontal movement in pan,
(ii) vertical movement or tilt, and (iii) change in depth-of-field by changing zoom.
In this paper, we limit ourselves to pan-only cameras, i.e., we assume that
a camera can move only in horizontal direction and its FoV is only described by its pan.
The pan of a camera is formally
defined using the following two parameters:
\begin{enumerate}
\item[(a)] ${R_s}$: Maximum coverage \emph{range} of the camera beyond which a target
can not be detected with acceptable accuracy in a binary detection test.
\item[(b)] $\theta$: The maximum sensing/coverage \emph{angle} of a camera on a certain direction.
This angle is also known as Angle of View (AoV).
\end{enumerate}

\begin{figure}
\centering
\includegraphics[width=0.4\textwidth]{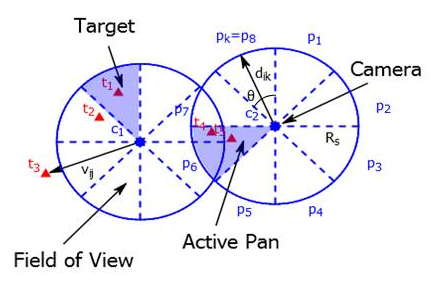}
\caption{Camera coverage parameters}
\label{fov}
\vskip -0.2in
\end{figure}

Thus, when a camera is oriented towards a particular direction, it can cover a circular sector (called a pan)
defined by ${R_s}$ and $\theta$. We assume that every camera possesses a specific number of non-overlapping pans,
of which, only one can be selected in a particular deployment. For example: a camera with FoV defined by
$\theta= \frac {\pi}{4}$ can pick any one of eight disjoint orientations.
Fig.~\ref{fov} depicts these parameters of camera coverage.
Here, two cameras $c_1$ and $c_2$ have eight pans each and can be oriented towards any of these
eight pans. We assume that cameras are homogeneous in terms of parameters.
Position of a target and a sensor are expressed through Cartesian coordinates
($x,y$) in a two-dimensional plane. $\overrightarrow{d_{ij}}$ is a unit vector which
cuts each pan (i.e., the sensing sector) into half representing the orientation of camera $c_i$ 
towards pan $p_j$.
$\overrightarrow{v_{it}}$ is a vector in the direction from camera $c_i$ to target $g_t$.

\noindent\textbf{Target in Sector (TIS) Test}:
With TIS test \cite{ai2006coverage}, one can verify whether a target $g_t$ is coverable by a given sensor $s_i$. 
To conduct this test, at first we calculate the angle $\phi_{it}$  between camera 
orientation $\overrightarrow{d_{ij}}$ of pan $p_j$ and the target vector $\overrightarrow{v_{it}}$.
\begin{equation}
\phi_{it} = cos^{-1}\left( \frac{\overrightarrow{d_{ij}}\cdot \overrightarrow {v_{it}}}{\mid \overrightarrow {d_{ij}}\mid \mid \overrightarrow {v_{it}}\mid} \right)
\end{equation}
A target is coverable by a camera's FoV if the span of its FoV contains the target and
the target is located within the sensing range of the camera.
Geometrically, $\overrightarrow{d_{ij}}$ divides the pan $p_j$ into two equal halves and if a target 
is located in either of them, it is coverable by that camera on the pan $p_j$.
Thus, the angle $\phi_{it}$ needs to be less than half of the AoV, i.e., 
$\phi_{it} \leq \frac {\theta}{2}$.
The other condition requires that the target has to be inside the maximum sensing range of the 
camera, i.e.,: $|v_{it}| \leq R_s$.

Conducting TIS tests over every pan $p_j$ of camera $s_i$ and every target $g_t$,
we can build a \emph{binary coverage matrix} $A^{M}_{N \times Q}$ of the network comprising of 
$M$ targets and $N$ cameras with $Q$ pans where an entry in the matrix can be calculated as:
\begin{equation}
a^{t}_{ij} = \left\{ \begin{array}{rl}
 1 &\mbox{ if target $g_t$ is covered by camera $s_i$ at pan $p_j$} \\
 0 &\mbox{ otherwise}
       \end{array} \right.
\end{equation}
\subsection{Fairness Index and Balancing  Index in k-coverage}\label{limitation_fairness}
Although the objective of the k-coverage problem is to cover each target at least $k$ times, 
it  may not be achievable if enough sensors are not 
available (i.e., in under-provisioned networks). 
Therefore, for under-provisioned networks a new metric needs to be defined to determine the 
superiority of one solution to another. The \emph{Fairness index} proposed by Jain  et 
al.~\cite{jain1998quantitative} has been traditionally used to measure \emph{fairness} in resource allocation 
systems. The visual sensor network can be thought of a similar resource allocation system 
where the sensors are the resources and the targets are the components of the system. 
Suppose in a particular configuration and orientation of sensors, the $m$ targets are  
covered $\psi_1, \psi_2, \ldots,\psi_m$ times respectively. Then the merit of this \emph{solution vector} 
$\left(\psi_1, \psi_2, \ldots,\psi_m\right)$ can be measured using the following equation:

\begin{equation}
\mathcal{F}\mathcal{I}  =  \dfrac{ (\sum_{t=1}^m \psi_t)^2 }{m \sum_{t=1}^m \psi_t^2}
\end{equation}
 
Thus, the fairness index of two solution vectors $(3,3,1,1)$ and $(2,2,2,2)$ each of 
which tries to cover four targets in a 3-coverage problem is 0.8 and 1.0 respectively 
which shows the solution $(2,2,2,2)$ is superior than $(3,3,1,1)$.
 
Even though fairness index is a good performance metric, solely focusing on maximizing 
it can result in reducing the total coverage. Fairness index usually 
identifies fairer solutions in the allocation of resources. Thus, in a 3-coverage problem with 
3 targets, (2, 2, 2) coverage is fairer than (2, 3, 2) although the second one is preferred 
because 3-coverage has not been attained at all in the solution (2, 2, 2) for any of the 
targets. 

Therefore, we modify the concept of fairness index to fit in to our purpose and introduce a 
more suitable metric called \emph{Balancing Index}, $\mathcal{B}\mathcal{I}$, which combines 
both fairness of coverage and maximization of the total coverage. 
It is defined as the product of fairness index and the ratio of achieved total coverage 
(i.e., $\sum_{t=1}^m \psi_t$) over total attainable coverage (i.e., $km$). Mathematically, 
the balancing index is:
\begin{equation}\label{eqn:bi}
\mathcal{B}\mathcal{I} = \mathcal{F}\mathcal{I} \times \dfrac{\sum_{t=1}^m \psi_t}{km} = \dfrac{(\sum_{t=1}^m {\psi_t})^2}{m \times \sum_{t=1}^m{\psi_t}^2} \times \dfrac{\sum_{t=1}^m \psi_t}{km}
\end{equation}
where $\psi_t$ is the number of sensors covering target $g_t$ and $m$ is the total number of targets. 
The balancing index value for (2, 2, 2) is:
$$\mathcal{B}\mathcal{I} = \dfrac{(2 + 2 + 2)^2}{3 \times (2^2 + 2^2 + 2^2)} \times \dfrac{(2 + 2 +2)}{3 \times 3} = 0.6666$$
The balancing index value for (2, 3, 2) is:
$$\mathcal{B}\mathcal{I} = \dfrac{(2 + 3 + 2)^2}{3 \times (2^2 + 3^2 + 2^2)} \times \dfrac{(2 + 3 + 2)}{3 \times 3} = 0.7472$$
The balancing index reflects that (2, 3, 2) coverage is better than (2, 2, 2) coverage in a 
3-coverage problem. Thus, it can be used as the performance metric. The higher the value of the 
balancing index, the better is the coverage.

\subsection{Problem Formulation}\label{prob_form}

The balanced $k$-coverage problem can formally be described as follows:

\emph{\textbf{Given:}} A set of targets, $\mathcal{T} = \{g_1, g_2, \ldots, g_m\}$ to be covered; 
a set of homogeneous directional sensors, $\mathcal{S} = \{s_1, s_2, \ldots, s_n\}$, 
each of which can be oriented in one direction of $q$ possible non-overlapping pans; the pan set, 
$\mathcal{P} = \{p_1, p_2, \ldots, p_q\}$. A collection of subsets, 
$\mathcal{F} = \{\Phi_{\langle i,j\rangle}| s_i \in \mathcal{S}, p_j \in \mathcal{P}\}$ 
can be generated based on a TIS test, where $\Phi_{ \langle i,j\rangle}$ is a subset of 
$\mathcal{T}$ and denotes the set of targets covered by selecting sensor $s_i$ and oriented in 
the direction of pan $p_j$.

\emph{\textbf{Problem:}} Find a sub-collection $\mathcal{Z}$ of $\mathcal{F}$, with the 
constraint that at most one $\Phi_{\langle i,j\rangle}$ can be chosen for the same 
sensor $s_i$, and the Balancing Index, $\mathcal{B}\mathcal{I}$ (defined in Equation~\ref{eqn:bi}), gets maximized.

\section{Generic k-coverage and its ILP formulation}\label{ILP}
In \cite{ai2006coverage}, the authors elaborate an ILP formulation 
to solve the maximization of \emph{single} coverage using minimum number 
of sensors (MCMS). The proposed ILP formulation can be easily 
extended for the $k$-coverage problem. In this section, at first we 
describe the necessary modifications and then we point out its 
shortcomings in providing a balanced solution for under-provisioned 
networks.

The parameters used for the formulation can be summarized as 
follows. $n$: the number of sensors; $m$: the number of targets; 
$q$: the number of orientation available for each directional sensor. 
The variables in the formulation are as follows.
$\psi_t$: an integer variable that has a value equal to the number of 
times a target $g_t$ is covered by directional sensors, limited up to 
a maximum value $k$; $\chi_{\langle i,j\rangle}$: a binary variable 
that has value one if the directional sensor $s_i$ uses the orientation 
$p_j$, and zero otherwise; $\xi_{t}$: an integer variable that counts 
the number of the directional sensors covering target $g_t$. 
$\Phi_{\langle i,j\rangle}$ is the set of targets that are covered by 
the sensor $s_i$ in its pan $p_j$. Using TIS, for each sensor $s_i$, 
incidence matrix $A^{i}_{(m \times q)}$ can be generated, where each 
of its elements would be:

\begin{equation}
a^{t}_{ij}= \left\{ 
	\begin{array}{lr}
		1 & t \in \Phi_{\langle i,j\rangle}\\
		0 & otherwise
	\end{array}
\right.
\end{equation}

Therefore, $\xi_t$ can be expressed as:\\
$$\xi_t = \sum\limits^{n}_{i=1}\sum\limits^{p}_{j=1}a^{t}_{ij}\chi_{\langle i,j\rangle}$$

Now, the ILP formulation for $k$-coverage problem becomes:
\begin{equation}\label{eq:objective}
	\textbf{maximize} \sum\limits_{t=1}^{m} \psi_t - \rho \sum\limits_{i=1}^{n}\sum\limits_{j=1}^{q} \chi_{\langle i,j\rangle}
\end{equation}	

subject to:	
	
\begin{equation}\label{eq:strt_cons}	
	\frac{\xi_t}{n} \leq \psi_t \leq \xi_t \quad \forall t = 1 \ldots m
\end{equation}	
\begin{equation}\label{eq:maximality_const}
		\psi_t \leq k
\end{equation}	
\begin{equation}\label{eqn:sum_over_all_pan}
	\sum\limits_{j=1}^{q}\chi_{\langle i,j\rangle} \leq 1\ \forall i = 1 \ldots n
\end{equation}
\begin{equation}\label{eq:end_cons}
	\chi_{\langle i,j\rangle} = 0\quad or\quad1 \quad \forall i = 1 \ldots n,\ \forall j = 1 \ldots q
\end{equation}
The objective function defined by Equation \ref{eq:objective} maximizes the coverage count 
of each target and imposes a penalty by multiplying
the number of sensors to be activated by a small positive penalty
coefficient $\rho$ $(\le1)$. There are $(m + np)$
variables and $(2m + n + np)$ constraints for the ILP. Equation
\ref{eq:strt_cons} represents a set of inequalities to indicate whether any
target $g_t$ is covered or not: if none of the sensors
cover target $g_t$, i.e., $\xi_t = 0$, then $\psi_t = 0$ to conform the right
inequality; if target $g_t$ is covered by at least one directional sensor, i.e.,
$\xi_t > 0$, since $\xi_t$ is bounded by $n$, $\xi_t/n$ 
is a real number less than one, then $\psi_t \geq 1$ to follow the left inequality. 
Constraints in Equation \ref{eq:maximality_const} make sure
that the coverage count of any target is bounded by $k$, i.e., even if a target is 
covered by more than $k$ times still the coverage
count will be considered as $k$, no additional benefit for covering a target more 
than $k$ times. Equation \ref{eqn:sum_over_all_pan} guarantees that one directional 
sensor has at most one orientation depending on whether it is activated or not.
\begin{figure}
\begin{center}
\begin{tabular}{c}
\resizebox{75mm}{!}{\includegraphics{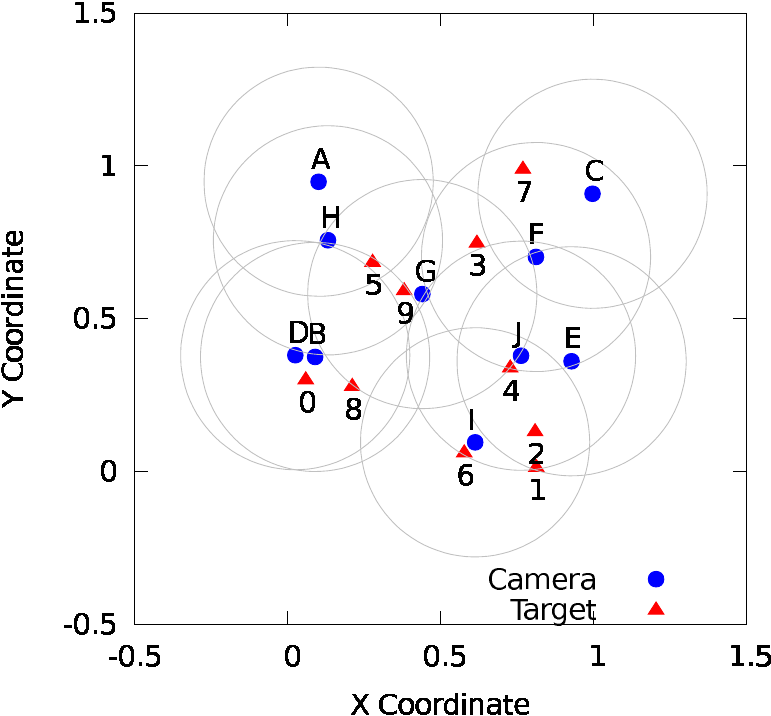}}\\
(a) Initial Configuration 10 Targets and 10 Cameras \\
\\
\resizebox{75mm}{!}{\includegraphics{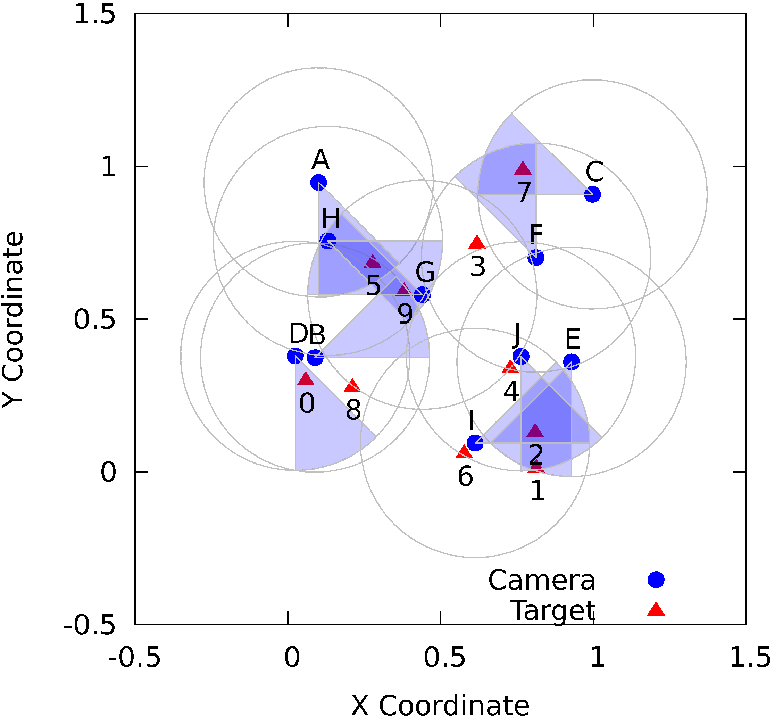}}\\
(b) Pan selection of each camera in ILP \\
\\
\resizebox{75mm}{!}{\includegraphics{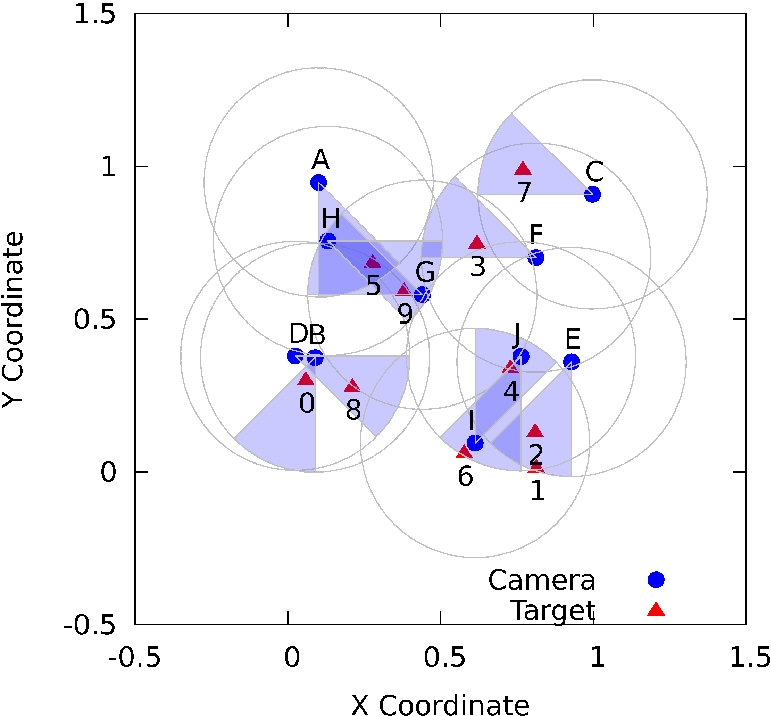}}\\
(c) Pan selection in balanced-optimal configuration \\
\end{tabular}
\caption{Illustrating imbalance coverage of ILP formulation}
\label{fig6}
\end{center}
\end{figure}

\noindent\textbf{Problem with generic ILP}:
The ILP formulation mentioned above for $k$-coverage problem does not focus on 
coverage balancing. To understand the problem, consider the 
scenario shown in Fig. \ref{fig6}(a).
Here we have shown a 3-coverage problem 
with 10 targets (red triangles) and 10 directional sensors 
(blue circles). FoV is defined by $\theta= \frac {\pi}{4}$. Table \ref{tab:analysis} summarizes 
the coverage counts under different conditions. 
Not all targets are 3-coverable:--it is possible to cover the targets $\{0,3,6,7\}$ at most twice.
Rest of the targets are at least 3-coverable. Clearly the network is under-provisioned.
After running the ILP, formulated above, we found a solution which is shown in Fig. \ref{fig6}(b).
The third column of Table \ref{tab:analysis} captures the coverage
achieved in ILP. Targets $\{2,5,9\}$ are covered thrice, targets $\{1,7\}$ are covered
twice, target $\{0\}$ is singly covered and noticeably targets $\{3,4,6,8\}$ are left 
totally uncovered. 

In summary, $40\%$ targets are not covered by any of the sensors
in the solution provided by the ILP.
Fig. \ref{fig6}(c) shows another possible solution of the same problem 
which we call \emph{balanced optimal} coverage (describe in Section~\ref{sec:balanced_optimal}). 
The fourth column of Table \ref{tab:analysis} shows the coverage
achieved in this new solution. Only target $\{5\}$ and targets $\{2,9\}$ are 3-covered
and 2-covered respectively however the rest of all targets are singly covered. 
Unlike ILP, none of the targets are left uncovered in this new solution. 
\begin {table}
\begin{center}
\caption {Detail analysis of example scenario in Fig. \ref{fig6}} \label{tab:analysis} 
\begin{tabular} {|c|c|c|c|}
\hline
Target ID & Maximum & Coverage & Coverage achievable \\[.5ex]
& possible coverage & achieved in ILP & in Balanced-optimal \\[.5ex]
\hline
0 &	2 &	1 &	1 \\[.5ex]
1 &	3 &	2 &	1 \\[.5ex]
2 &	3 &	3 &	1 \\[.5ex]
3 &	2 &	0 &	1 \\[.5ex]
4 &	5 &	0 &	2 \\[.5ex]
5 &	4 &	3 &	3 \\[.5ex]
6 &	2 &	0 &	1 \\[.5ex]
7 &	2 &	2 &	1 \\[.5ex]
8 &	3 &	0 &	1 \\[.5ex]
9 &	3 &	3 &	2 \\[.5ex]
\hline
\end{tabular}
\end{center}
\vskip -0.2in
\end{table}

\section{Balanced k-coverage} \label{balanced_k}
\begin{figure}
        \centering        
        \includegraphics[width=.25\textwidth]{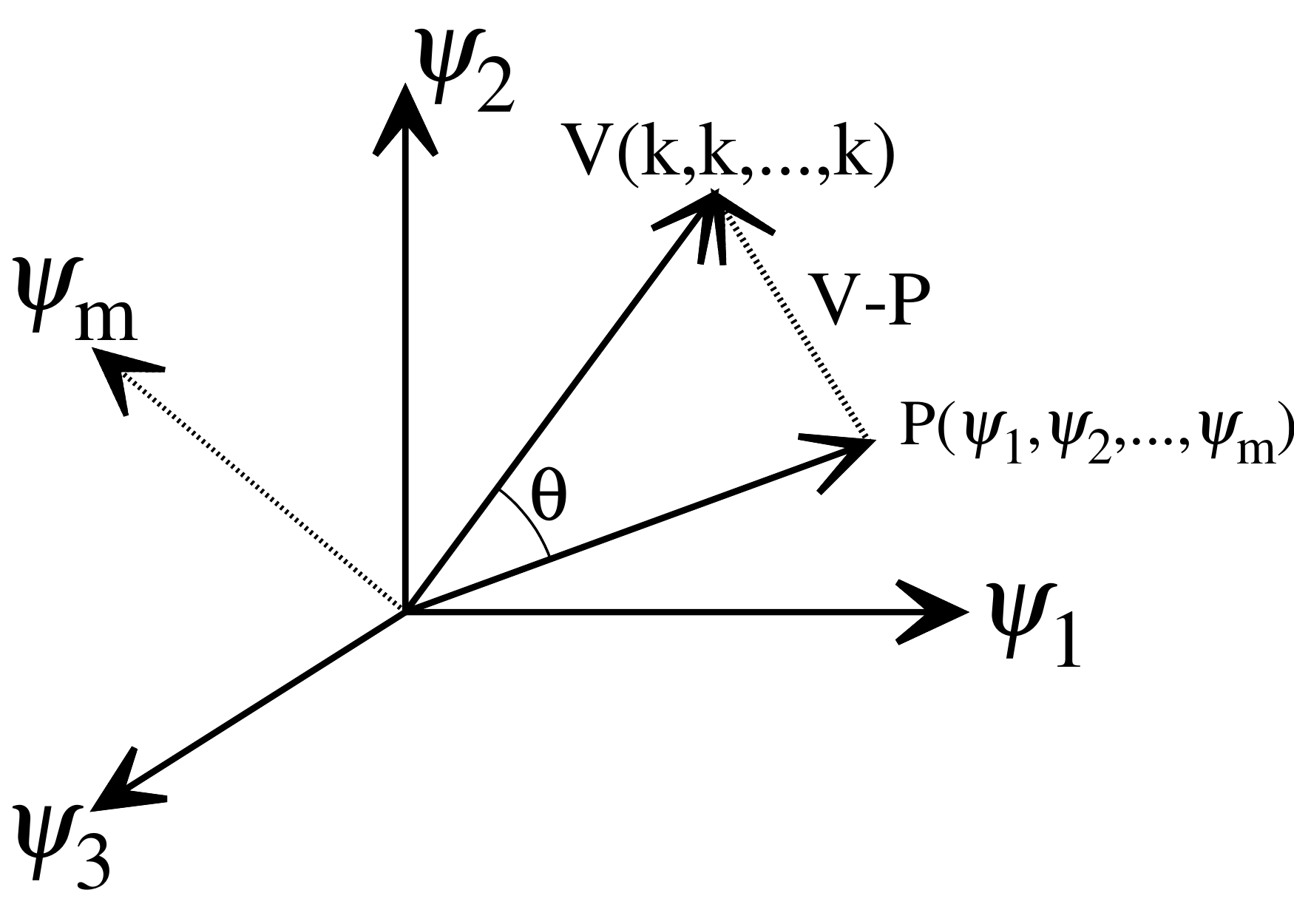}
        \caption{Coverage displayed in m-dimensional space}
	\label{fig:vector}
\vskip -0.2in
\end{figure}

In order to improve the balancing of coverage, we need to modify the ILP formulation,
specially its objective function. In particular, the objective function concurrently needs to 
keep track of balancing of coverage while pursuing $k$-coverage.
To keep the problem tractable, we may consider the solutions of $k$-coverage
problem as vectors in an $m$-dimensional vector space.
The coverage counts of targets can be considered as 
individual dimensions in the $m$-dimensional space.
Fig. \ref{fig:vector} shows this typical scenario.
Each axis represents the coverage count of each target 
and there is a total of $m$ possible targets. 
The vector ${\bf V}$ $\equiv$ $(k, k, \ldots, k)$ represents 
the desired solution vector.
Let us consider the vector ${\bf P}$ which 
represents the achieved coverage by an arbitrary algorithm. 
Hence vector ${\bf P}$ can be represented as ${\bf P}$ $\equiv$ $(\psi_1, 
\psi_2, \ldots, \psi_k)$.
Although it is highly desirable to align the vector ${\bf P}$ with the 
vector ${\bf V}$, but practically it may not be achieved by an algorithm.
In such cases, the goal should be  
to \emph{minimize} the \emph{distance} between these two vectors, 
$\overrightarrow{{\bf PV}}$ [$\overrightarrow{{\bf PV}} = {\bf V} - {\bf P}$]
or to \emph{minimize} the \emph{angle} between them (i.e., the $\theta$). We 
formally describe both of these intuitive approaches below.
\subsection{Minimizing the vector distance}\label{min_dis}
The vector distance between the actual coverage vector ${\bf P}(\psi_1, 
\psi_2, \ldots, \psi_k)$ and the expected coverage vector ${\bf V}(k, k, \ldots, k)$ can be 
calculated as follows: 
$$d({\bf V} , {\bf P}) = ||{\bf V} - {\bf P}|| = \sqrt{\sum_{t=1}^m (k- \psi_t)^2}$$
where $k$ is the number of times the targets are to be covered and $\psi_t$ is the 
achieved coverage of target $g_t$. The minimization problem
remains the same even if we ignore the square root.
Thus, the ILP formulation described in Section \ref{ILP} 
can be easily modified to achieve the goal. 
We can simply modify the objective function (Equation \ref{eq:objective}) and incorporate the square of the vector distance
leaving all constraints (Equations \ref{eq:strt_cons} - \ref{eq:end_cons}) unchanged. 
The new objective function will be as follows:
$$\textbf{minimize} \quad \sum\limits_{t=1}^{m} (k -\psi_{t})^2 + \rho \sum\limits_{i=1}^{n}\sum\limits_{j=1}^{q} \chi_{\langle i,j\rangle}$$
The modified objective function is no more linear but quadratic in nature. Therefore,
the new formulation is an \emph{integer quadratic programming} problem or IQP in short.
\subsection{Minimizing the angle}
Another approach is to minimize the angle between the two vectors ${\bf V}$ and ${\bf P}$ using the following 
equation:
$$\theta = cos^{-1}\left(\frac{V.P}{||V||||P||}\right) = cos^{-1}\left(\frac{\sum_{t=1}^m\psi_t}{\sqrt{m \sum_{t=1}^m\psi^2_t}}\right)$$
However, minimizing the angle between vectors can not differentiate isomorphic solutions
like $(1,1,\ldots,1)$, $(2,2,\ldots,2)$, $(3,3,\ldots,3)$  ... etc. because all of these
vectors make an angle zero with the ideal solution vector $(k,k,\ldots,k)$.
Therefore, we do not explore this issue further in this paper.
\subsection{Maximizing the balancing index} \label{sec:balanced_optimal}
Finally, the objective function of the ILP
can be modified to incorporate {\bf balancing index} as defined in Section~\ref{limitation_fairness}
and achieve true coverage balancing. The necessary modification is as follows:
$$\textbf{maximize} \quad \frac{1}{km^2}\times\frac{ (\sum_{t=1}^{m} \psi_{t})^3 }{\sum_{t=1}^{m} \psi_{t}^2} -\rho \sum\limits_{i=1}^{n}\sum\limits_{j=1}^{p} \chi_{\langle i,j\rangle}$$	
subjected to the constraints described in Equations \ref{eq:strt_cons} - \ref{eq:end_cons}.
Note that, the new objective function is non-linear in nature. Therefore, the modified 
formulation falls within the domain of 
\emph{integer non-linear programming} problems or INLP in short.

\begin{algorithm}
\caption{Centralized Greedy $k$-coverage Algorithm (CG$k$CA)}
\label{k_algo}
\begin{algorithmic}[1]

\renewcommand{\algorithmicrequire}{\textbf{Input:}}
\renewcommand{\algorithmicensure}{\textbf{Output:}}
\Require{$\Phi_{\langle i,j\rangle}$ \{set of targets covered by sensor $s_i$ in pan $p_j$\}}
\Ensure{$\mathcal{Z}$} \space \{a collection of $\langle$active sensor, orientation$\rangle$ pairs\}
\State $\mathcal{Z} \gets \emptyset$
\State $\mathcal{Y} \gets$ \{set of inactive nodes\}
\State $\mathcal{T} \gets \{g_1, g_2, \ldots g_m\}$ \{set of all targets\}
\State $\mathcal{C} \gets \{c_1, c_2, \ldots c_m\}$ \{$c_i$ is the coverage count of $g_t$\}

\Repeat
\State $maxincentive \gets 0$
\State{$\alpha \gets 0$ for linear benefit; $\alpha \gets 1$ for quadratic benefit}
\For{$\forall i \in \mathcal{Y}$}
	\For{$\forall j \in \mathcal{P}$}
		\State $incentive \gets$ \Call{Benefit}{$\Phi_{\langle i,j\rangle},\mathcal{Z},\alpha$}	\label{step:benefit}	
		\If{$incentive > maxincentive$}
			\State $maxincentive \gets incentive$
			\State $\langle i^{max},j^{max}\rangle = \langle i,j\rangle$
		\EndIf
	\EndFor
\EndFor

\State $\mathcal{Z} \gets \mathcal{Z} \cup \langle i^{max},j^{max}\rangle$
\State $\mathcal{Y} \gets \mathcal{Y} \backslash \{i^{max}\}$

\Until {$maxincentive = 0$}
\State\Return $\mathcal{Z}$
\end{algorithmic}
\end{algorithm}

\section{Balanced $k$-Coverage heuristics}\label{heuristics}
The ILP, IQP, and INLP formulation of the problem mentioned in the previous section 
can be used to find the optimal solutions, however, 
they are not scalable in large problem instances. 
Therefore, we present Centralized Greedy 
$k$-coverage Algorithm (CG$k$CA), a polynomial time greedy heuristic that would 
converge faster and by suitably choosing appropriate set of sensors would also balance the coverage. 
CG$k$CA uses a \emph{benefit function}, which calculates the incentive of selecting a 
particular $\langle$sensor, pan$\rangle$ pair at each step.

The basic idea of CG$k$CA is to greedily choose and activate the $\langle$sensor, pan$\rangle$ 
pair which provides the maximum benefit. In each iteration, the incentives 
of all inactive $\langle$sensor, pan$\rangle$ pairs are calculated using 
the benefit function. The pair with maximum incentive is selected and the sensor is 
activated towards the corresponding pan. Ties are broken 
arbitrarily by choosing among the pairs providing maximum incentive. The 
algorithm terminates when all the sensors 
are activated or when all the targets are at least $k$-covered. 
The pseudo code of CG$k$CA is given in Algorithm \ref{k_algo}. 
\begin{algorithm}
\caption{Benefit Function for $k$-coverage}
\label{benefit}
\begin{algorithmic}[1]
\renewcommand{\algorithmicrequire}{\textbf{Input:}}
\renewcommand{\algorithmicensure}{\textbf{Output:}}
\Require{$\Phi_{\langle i,j\rangle}$ \{set of targets covered by sensor $s_i$ in pan $p_j$\}, $\mathcal{Z}$ \{set of assigned $\langle$active sensor, orientation$\rangle$ pairs\}}, $\alpha$ \{$\alpha = 0$ for linear benefit and $\alpha = 1$ for quadratic benefit\}
\Ensure{$incentive$ \{an integer containing the total incentive for $\langle i,j\rangle$ for given $\mathcal{Z}$\}}
\Function{Benefit}{$\Phi_{\langle i,j\rangle}, \mathcal{Z},\alpha$}
\State $incentive \gets 0$

\For{$\forall t \in \Phi_{\langle i,j\rangle}$}
	\State Calculate the coverage $c_t$ of target $t$ using $\mathcal{Z}$
	\If{$c_t < k$}
		\If{$\alpha = 0$}	
			\State	$increment \gets 1$
		\ElsIf{$\alpha = 1$} 
			\State $increment \gets (k - c_t)^2 - (k - c_t -1)^2$
		\EndIf
	\State	$incentive \gets incentive + increment$
	\EndIf
\EndFor
\EndFunction
\State\Return $incentive$ 
\end{algorithmic}
\end{algorithm}

\noindent{\bf Benefit function}: In order to choose from all inactive sensor-pan pairs, 
CG$k$CA makes a call to a benefit function in step \ref{step:benefit} of 
Algorithm \ref{k_algo}. The benefit function, which is defined in Algorithm~\ref{benefit}, 
calculates two different kinds of benefit:--(i)
linear benefit and (ii) quadratic benefit. The three parameters of the benefit function 
is as follows:

$\Phi_{\langle i,j\rangle}$: set of targets covered by sensor-pan pairs $\left<i,j\right>$.

$\mathcal{Z}$: set of sensor-pan pairs activated so far by the greedy algorithm before this step.

$\alpha$: a boolean parameter to indicate whether the benefit function should calculate either linear 
or quadratic benefit. $\alpha = 0$ means linear and quadratic otherwise.
\begin {table}[b]
\begin{center}
\caption {Incentive table for 3-coverage problem ($k=3$)} \label{tab:incentive} 
\begin{tabular} {|c|c|c|}
\hline
Coverage count of a & Incentive in linear & Incentive in quadratic \\[.5ex]
target at this stage & benefit function & benefit function \\[.5ex]
\hline
$c_t$ & $min\{1,k-c_t\}$ & $(k-c_t)^2 - (k-c_t-1)^2$ \\[.5ex]
\hline
0 & 1 &  5\\[.5ex]
\hline
1 & 1 & 3 \\[.5ex]
\hline
2 & 1 & 1 \\[.5ex]
\hline
\end{tabular}
\end{center}
\vskip -0.2in
\end{table}
The linear benefit is calculated as follows. For each target $g_t$, let us define a variable $c_t$ which
assumes a value equal to the coverage count of that target by the set of sensor-pan pairs in set $\mathcal{Z}$, 
i.e.: 
$$c_t = \text{number of sensing regions in } \mathcal{Z} \text{ that cover target } g_t$$

Then the total linear benefit of activating sensor-pan pair $\left<i,j\right>$ at this stage of CG$k$CA is: 
$$\sum_{t \in \Phi_{\langle i,j\rangle} \land c_t < k}min\{1,k-c_t\}$$
Note that, the linear benefit function ignores the coverage count of a target at this 
stage of the greedy algorithm and assigns same incentive for covering a target 
irrespective of its coverage count.
The quadratic benefit function eliminates this drawback
and provides more incentive for covering a less covered target as opposed to highly covered 
targets. The incentive of covering a target is quadratic in nature and is defined by 
$ k^2 - (k-i)^2$ for a target that is covered $i$ times. 
Thus the quadratic benefit of activating sensor-pan pair $\left<i,j\right>$ at this stage 
of CG$k$CA becomes: 
$$\sum_{t \in \Phi_{\langle i,j\rangle} \land c_t < k}(k-c_t)^2 - (k-c_t-1)^2$$
Table \ref{tab:incentive} shows the incentive values for both linear and quadratic benefit functions
that will be rewarded while solving  a 3-coverage problem. Algorithm \ref{benefit} provides the complete 
pseudo-code of the benefit function. For rest of the paper, when the greedy algorithm runs with the linear benefit function, 
the approach is dubbed as \emph{Greedy Linear}. Similarly, when it runs with quadratic benefit function, we call it 
\emph{Greedy Quadratic}.
\noindent{\bf Time complexity}:
CG$k$CA needs the incidence matrix as input. To generate the incidence matrix, 
we need to iterate over all the $n$ sensors in each of their $q$ pans and check for 
each target if they satisfy the TIS test. As the TIS test takes $O(1)$ time for a 
specific sensor in its specific pan and a fixed target, the whole generation 
would take $O(nmq)$ time.

The major contributor to time complexity of the algorithm is due to 
the calculation of maximum incentive. The loop from line 5 to 20 is 
executed at most n times to check for each sensor. In each iteration, the 
benefit function will be called $O(nq)$ times. On each call, benefit function 
will check all the target within that pan and calculate the coverage counts of 
those targets and then the incentive of the $\langle$sensor, pan$\rangle$ pair. 
It will cost $O(mn)$. Therefore, the cost of each iteration is $O(n^2qm)$. 
Thus, the overall time complexity of the algorithm  becomes $O(n^3qm)$.

\section{Experimental Results}\label{results}
In order to verify and compare the effectiveness of proposed ILP, IQP, and INLP formulations and
the greedy algorithm with linear and quadratic benefit functions, we perform rigorous
simulation experiments. We use \emph{balancing index}, $\mathcal{B}\mathcal{I}$, defined in 
Section~\ref{limitation_fairness} as the performance metrics of comparison. A higher
value of $\mathcal{B}\mathcal{I}$ indicates highly balanced coverage. At the end,  
we also comment on the sensor usage by the different formulations and greedy heuristics.

\subsection{Simulation Environment}
The deployment area is modelled as a 2D grid where targets are considered  
points in the grid and the sensors are modelled as directional sensors with FoV, $\theta= \frac {\pi}{4}$. 
We run two different types of experiments. In one kind, we keep the number of sensors 
fixed at 50 and vary the number of targets from 5 to 125 and in another kind, we keep 
the number of targets fixed at 50 and vary the number of available sensors from 
20 to 115. For both scenarios the sensing range $R_{s} = 25$ units and the grid size is 
$125 \times 125$ sq. units. 
In both cases, the scenarios are generated in such a way that the smaller scenario 
is a subset of a larger scenario. This ensures a consistent evaluation of the impact of the
enlarged population of sensors/targets by retaining all the
``features'' of the previous environment and simply making it
better/worse. 

\subsection{Performance comparison of different approaches}
We capture performance of the proposed approaches by measuring 
$\mathcal{B}\mathcal{I}$ while changing the network's state from 
under-provisioned to over-provisioned type and vice-versa.
\begin{figure}
        \centering
		\begin{subfigure}{.4\textwidth}
			\centering        
     		\includegraphics[height=0.97\textwidth, width=1.1\textwidth]{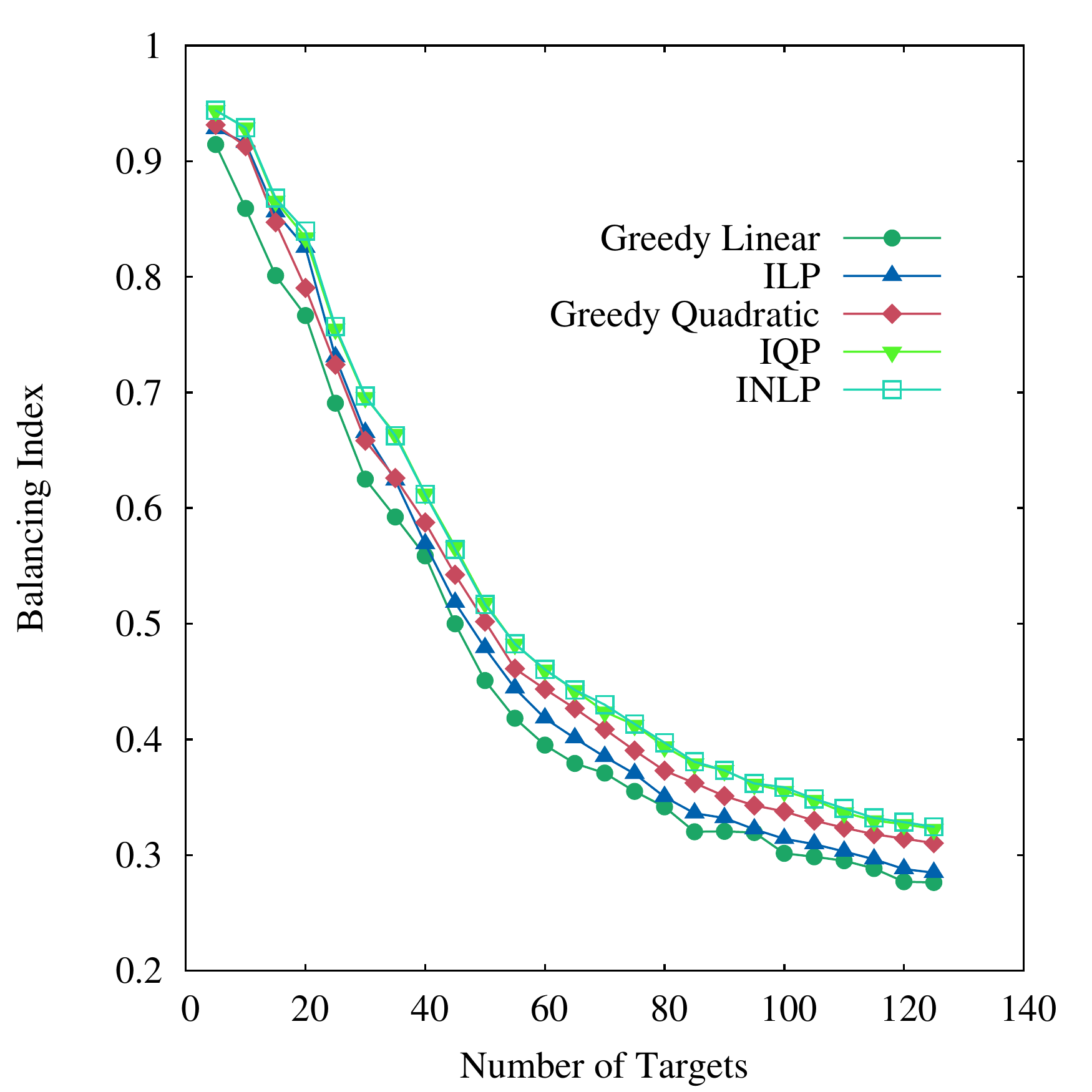}
        	\caption{Performance comparison while varying the number of targets; the number of sensors is fixed at 50}
	        \label{fig:all_div_50}
	    \end{subfigure}    
        %
        \begin{subfigure}{.4\textwidth}
			\centering        
     		\includegraphics[height=0.97\textwidth, width=1.1\textwidth]{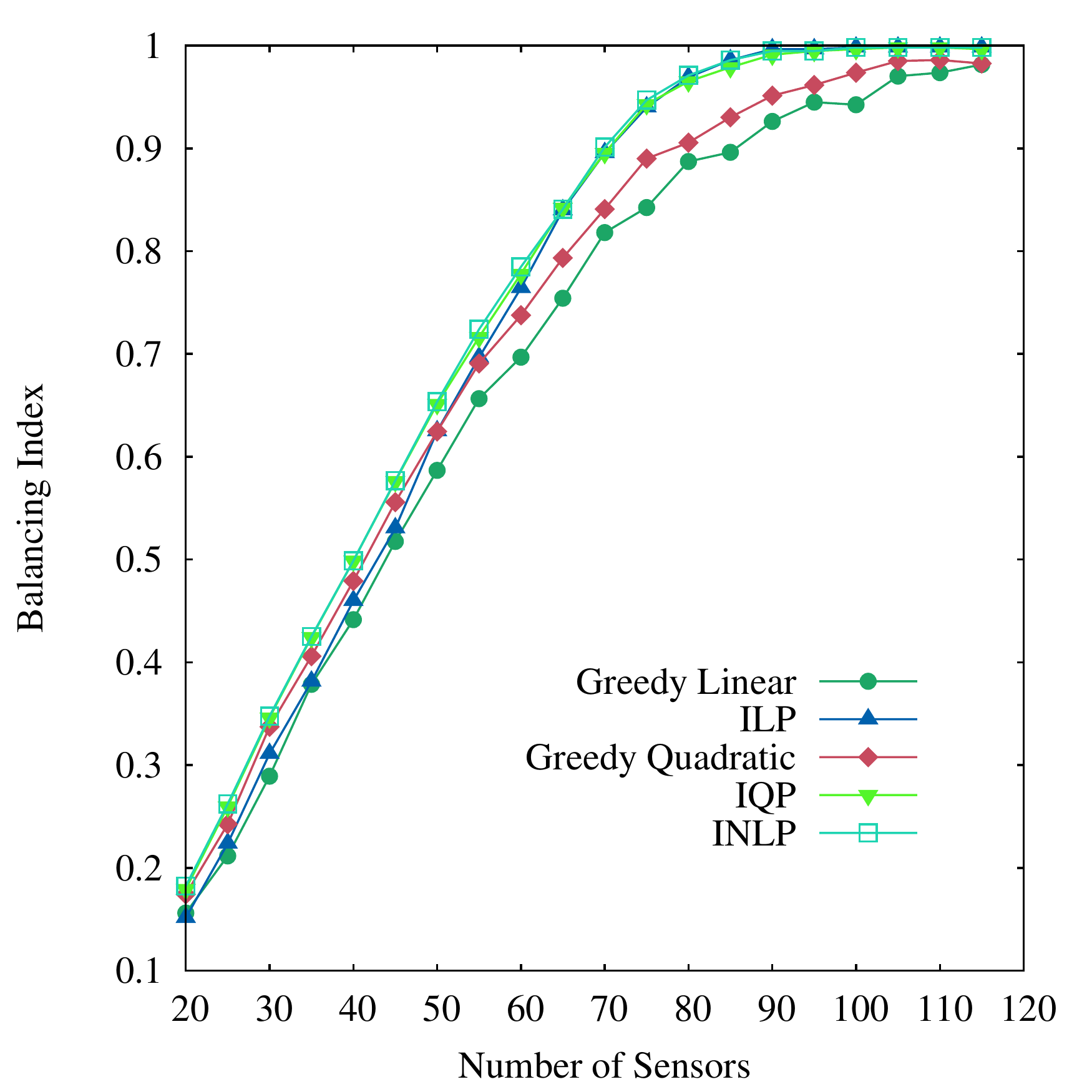}
        	\caption{Performance comparison while varying the number of sensors, the number of targets is fixed at 50}
	        \label{fig:all_div_t50}
	    \end{subfigure}
	    \caption{Effect on Balancing Index}\label{fig:all_div}
\vskip -0.2in
\end{figure}

In the first type of environment, when we increase the number of targets gradually from 5 to 125
keeping number of sensors fixed at 50, the network's state changes from over-provisioned
to under-provisioned as the number of targets slowly overwhelms the number of sensors. 
As a result, from Fig. \ref{fig:all_div_50} it is clearly evident that the curves move 
farther away from the ideal coverage ($\mathcal{B}\mathcal{I} =1$). 
In the whole downward progress, INLP and IQP clearly outperform all other methods. 

In the second type of environment, increasing the 
number of available sensors and keeping the number of targets fixed at 50
shifts the state of the network from under-provisioned
to over-provisioned. This behaviour is reflected in all curves as there is an upward 
movement toward the ideal coverage in Fig. \ref{fig:all_div_t50}. 
Also here, the performance of INLP and IQP formulations exceeds 
the other approaches.

In both cases, the Greedy Quadratic shows greater coverage and balancing 
than the ILP formulation in under-provisioned condition, however when the network 
starts to contain larger number of sensors relative to the number of targets, 
ILP crosses the Greedy Quadratic curve. In a completely over-provisioned scenario, 
ILP almost merges with IQP and INLP formulations. Greedy Linear fails to keep up 
with all other formulations in all cases. One notable point is that, IQP curve is 
almost merged with INLP curves throughout the whole simulation but never exceeds 
the performance of INLP. We can conclude that the Greedy Quadratic approximates 
the optimal behaviour very closely and with a very reasonable amount of 
computational time.

\begin{figure}
        \centering
			\centering        
     		\includegraphics[width=0.5\textwidth]{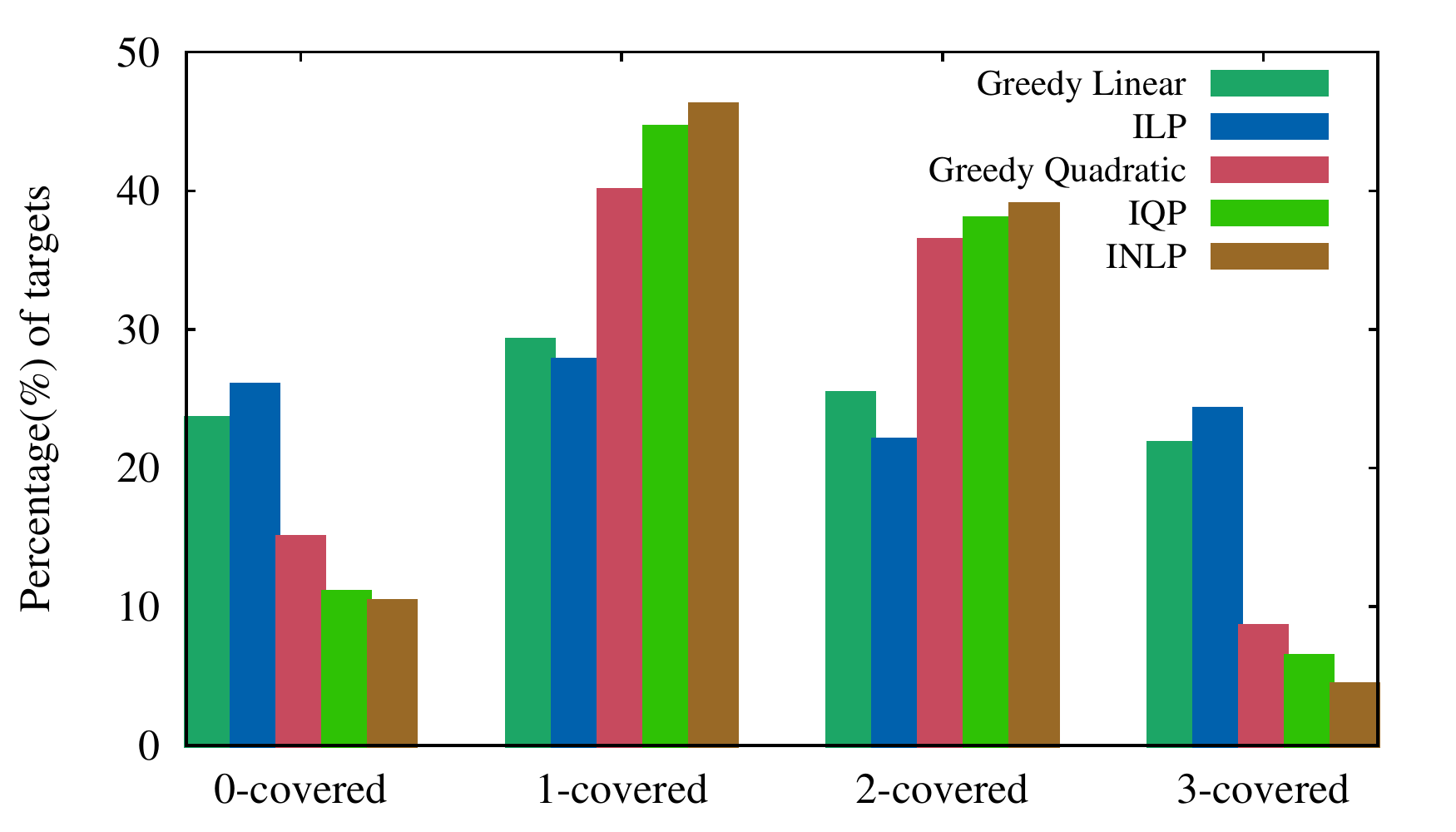}
        	Number of sensors = 50, number of targets = 100
	        \label{fig:under_50}
        %
	    \caption{Coverages in Under-provisioned Network}\label{fig:under}
\vskip -0.2in
\end{figure}

\subsection{Coverage Analysis}
Now we present a detailed coverage analysis of both under-provisioned and over-provisioned
networks.
\subsubsection{Under-provisioned Networks}
We have deliberately created an under-provisioned network 
by keeping the number of targets to 100 and the number of sensors to 50.  
Clearly there are not enough sensors available to cover all the targets 
$k$ times ($k=3$). The detailed coverage ananlysis of such network is shown 
in Fig.~\ref{fig:under}. In this under-provisioned network, INLP as well as IQP tries to reduce 
the number of uncovered targets at the expense of number of 
higher covered targets. The percentage of uncovered targets 
is 10.4\% and 11\% for INLP and IQP respectively. 
The Greedy Quadratic roughly approximates this behaviour 
by reducing the number of uncovered targets to 15\%. 
The other two approaches (ILP and Greedy Linear) do not 
focus on coverage balancing and as a result they increase 
the coverages of some targets, keeping a large number of 
targets totally uncovered.

\subsubsection{Over-provisioned Network}
\begin{figure}
        \centering
			\centering        
     		\includegraphics[width=0.5\textwidth]{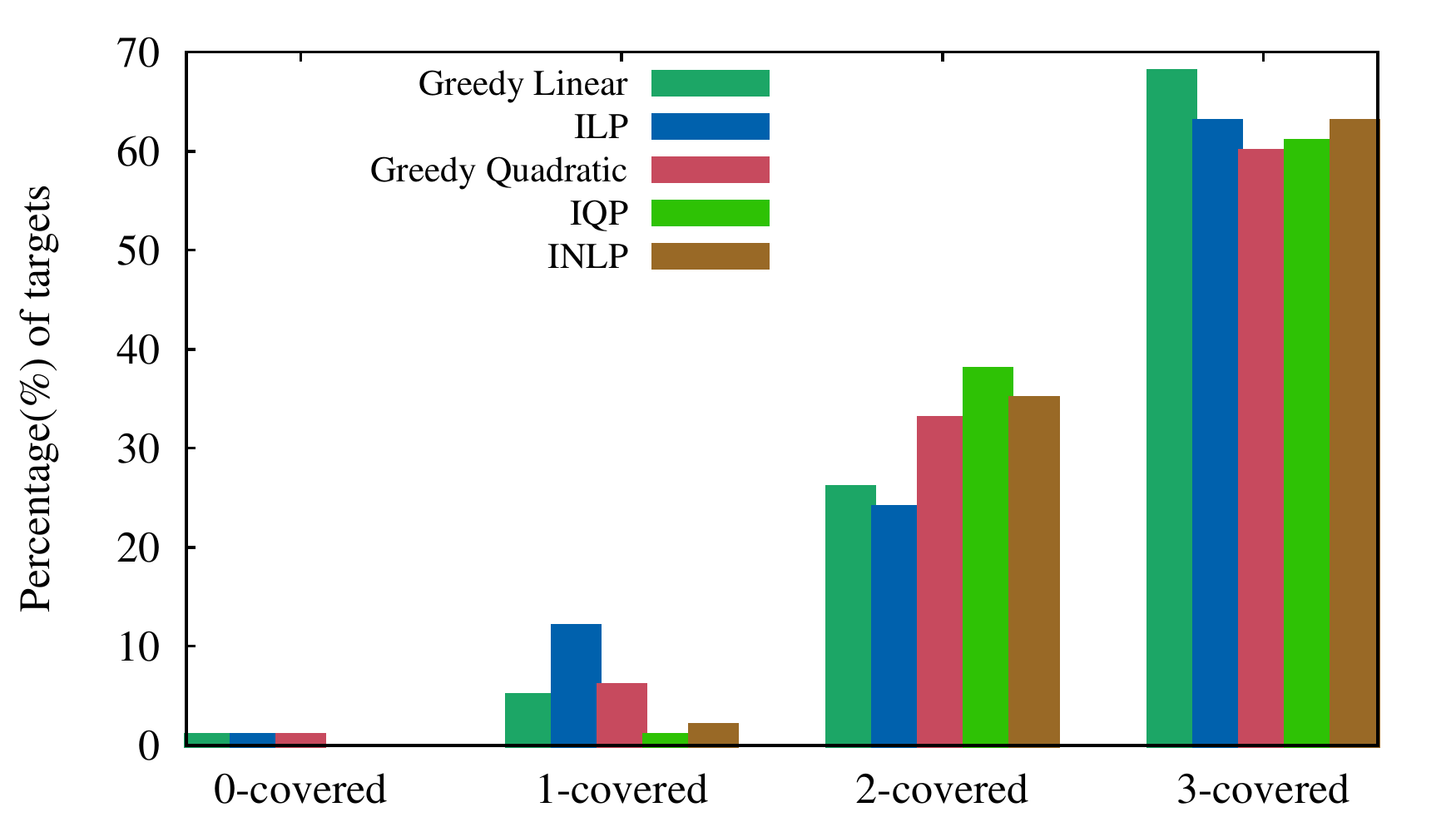}
        	Number of sensors = 50, number of targets = 20
	        \label{fig:over_50}
        %
	    \caption{Coverages in Over-provisioned Network}\label{fig:over}
\vskip -0.2in
\end{figure}

Next, we have created an over-provisioned network with 50 sensors and 20 targets. 
Clearly this is an over-provisioned network
since there are enough available sensors to 
cover most of the targets at least $k$ times. Again the detailed coverage analysis of such network
is shown in Fig. \ref{fig:over}.
INLP and IQP formulations again gives more importance to 
targets covered less number of times. 
As a consequence, there is no uncovered targets for INLP and IQP formulations. 
Due to abundance of available sensors, all the 
formulations tries to increase the number of 2-covered and 3-covered 
targets and INLP and IQP formulations exceeds the other approaches in doing so. 
Among the greedy algorithms, the Greedy Quadratic 
performs much better as it reduces the number of lower
covered targets.
\subsection{Sensor usage Analysis}
\begin{figure}
        \centering
		\begin{subfigure}{.4\textwidth}
			\centering        
     		\includegraphics[width=\textwidth]{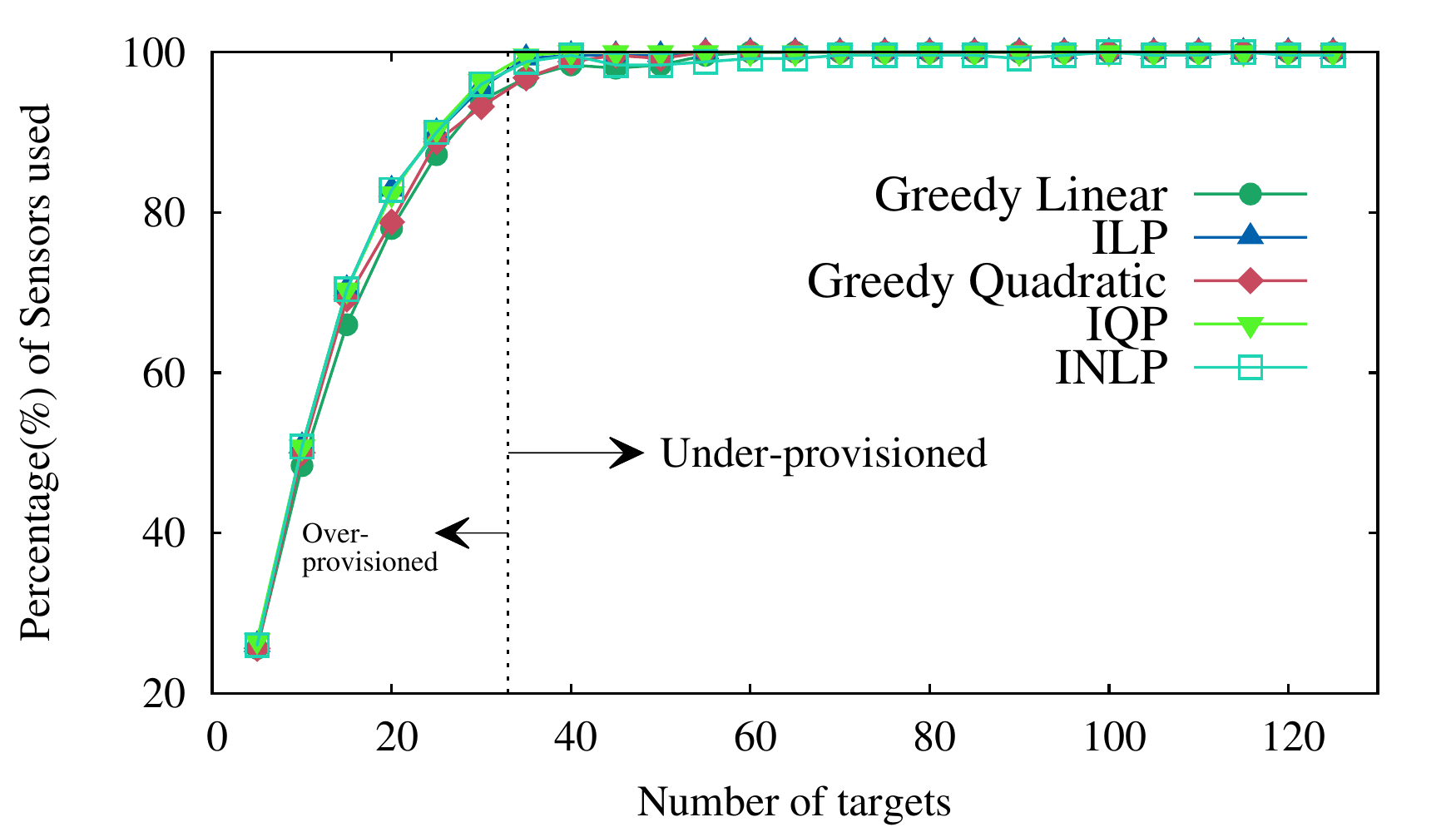}
			Number of sensors fixed at 50
        	\caption{Sensor usage while varying number of targets}
	        \label{fig:sensors_used_50}
	    \end{subfigure}    
		\hfill
        \begin{subfigure}{.4\textwidth}
			\centering        
     		\includegraphics[width=\textwidth]{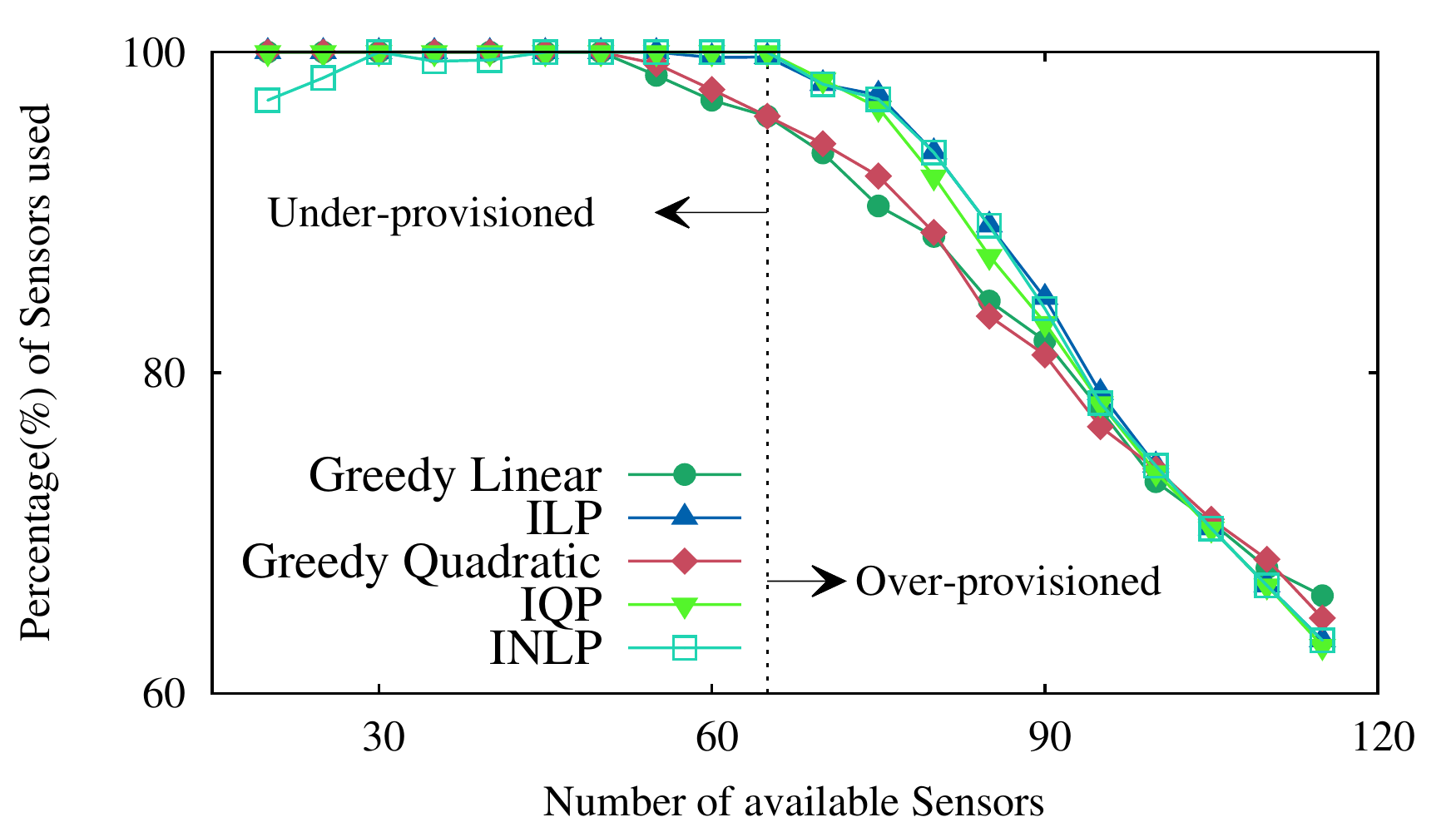}
			 Number of targets fixed at 50
        	\caption{Sensor usage while varying number of sensors}
	        \label{fig:sensors_used_t50}
	    \end{subfigure}
	    \caption{Analysis of Sensors usage}\label{fig:sensors_usage}
\vskip -0.2in
\end{figure}

Fig. \ref{fig:sensors_used_50} shows the percentage of sensor usage for the 
scenarios of Fig. \ref{fig:all_div_50}. Percentage of sensors 
used gradually increases  with the number of targets until all the cameras 
are activated. All the formulations perform almost similarly. 
Fig. \ref{fig:sensors_used_t50} is the sensor 
usage diagram for scenarios in Fig. \ref{fig:all_div_t50}. As the number of sensors 
gradually increases, lesser percentage of available sensors are activated.  
Although the number of sensors used by all formulations are similar when the
number of available sensors is lower, the situation changes with the increase in
available sensors. 
With 50 to 115 sensors, sensor usage of 
Greedy Linear and Greedy Quadratic  is less than others. 
However in terms of coverage, they were always outperformed 
by IQP and INLP.

Interestingly, both Fig. \ref{fig:sensors_used_50} and Fig. \ref{fig:sensors_used_t50}
shows a clear transition from under-provisioned network to over-provisioned network
and the sensor usage phenomena also changes accordingly.
In Fig. \ref{fig:sensors_used_50} when the number of targets exceeds
50 and in Fig. \ref{fig:sensors_used_t50} when the number 
of sensors falls below $55$, the networks become under-provisioned.
We can see that all of our formulations used up almost all the 
sensors in such under-provisioned networks. As a result all the 
curves become almost linear and parallel to the x-axis in these regions.
Reducing sensor usage only happened in the over-provisioned networks 
(i.e., the other side of the plots). 
In this region, although all the formulations used almost the same 
number of sensors as shown in Fig. \ref{fig:sensors_used_50}, 
but INLP and IQP had much better coverage balancing over other formulations
(see Fig. \ref{fig:all_div_50}). Among the variants of the greedy algorithm,
Greedy Quadratic shows better coverage balancing over Greedy Linear, 
although they use the same number of sensors. From Fig. \ref{fig:sensors_used_t50} 
and Fig. \ref{fig:all_div_t50}, we can draw a similar conclusion.

\section{Related Works}\label{relate}
A large volume of research exists for the $k$-coverage problem 
where the researchers worked only with omnidirectional sensors
~\cite{yen2006expected}~\cite{hefeeda2007randomized}~\cite{ammari2010study}~\cite{kumar2004k}.
Li-Hsing Yen et al.~\cite{yen2006expected} formulated an exact mathematical 
expression for the expected area that would be $k$-covered. 
Hafeeda and Bagheri~\cite{hefeeda2007randomized} modelled $k$-coverage problem 
as optimal hitting set problem which is NP-hard \cite{garey2002computers}. 
Their proposed~\cite{hefeeda2007randomized} $k$-coverage algorithm is inspired by 
the approximation algorithm in~\cite{bronnimann1995almost} and they suggested both centralized 
randomized $k$-coverage and distributed randomized $k$-coverage algorithms. 
In \cite{ammari2010study}, the authors has worked with connectivity and coverage of WSN in 
3D space. They proved that if there are $k$ sensors with spherical sensing range 
in a Reuleaux tetrahedron then all the targets in that Reuleaux tetrahedron 
will be $k$-covered. Most of the works discussed so far assume that the locations are known. 
The algorithm presented by Yigal Bejerano in \cite{bejerano2008simple} can efficiently 
verify the $k$-coverage without any prior location information.

In \cite{tian2002coverage}, Tian and Georganas 
devised a node scheduling algorithm which would turn off redundant sensors 
without reducing the overall coverage of the network. This would be achieved 
by turning off only those sensors whose coverage region is covered by its 
neighbouring active sensors. In \cite{kumar2004k}, Kumar et al. provided a solution to 
$k$-coverage problem where at any point of time most of the nodes are in sleep state. 
The authors study the lifetime of a network using different types of node distributions
such as $\sqrt{n} \times \sqrt{n}$ grid model, random uniform distribution and Poisson distribution.

There exist quite a few works using directional sensors. Jing and 
Abouzeid~\cite{ai2006coverage} formulate the coverage problem 
using directional sensors as Maximum Coverage using Minimum Sensors (MCMS)
problem, and provide both centralized and decentralized greedy solutions. 
In \cite{munishwar2013coverage}, Munishwar and Abu-Ghazaleh present new algorithm which improves 
the greedy approaches of \cite{ai2006coverage}. The 
optimal solution to $k$-coverage problem has been proved to be NP-hard by 
Fusco and Gupta in \cite{fusco2009selection}. They also modelled the sensors 
to have a fixed viewing angle and overlapping pans. However, all of these works overlooked the 
coverage imbalance issue.

\section{Conclusions and Future Works}\label{conclusion}
Our work addresses a novel problem of coverage imbalance 
in $k$-coverage of VSN. 
Coverage imbalance is a serious problem in under-provisioned 
networks, where the networks does not have enough sensors to ensure $k$-coverage
of all the targets.
We extended the traditional ILP designed for single coverage 
and applied it to solve the multi-coverage problem.
However, the extension does not balance the coverage of the targets.
We further designed quadratic (IQP) and non-linear (INLP) version of the ILP 
that are capable of addressing the coverage balancing.
IQP minimizes the vector distance between the attained and expected coverage.
INLP maximizes the Balancing Index, $\mathcal{B}\mathcal{I}$, which is the 
product of Fairness Index ($\mathcal{F}\mathcal{I}$) and average coverage.
As ILP, IQP, and INLP are not scalable for large problem instances, 
we developed a greedy approach, CG$k$CA with two variants of incentive mechanism
namely Greedy Linear and Greedy Quadratic.
Even though both greedy approaches are outperformed by optimal algorithms, 
in under-provisioned networks Greedy Quadratic closely approximates the optimal solutions.
We ran computer simulations to verify efficacy of the proposed formulations. 
In future, we plan to carry out more simulations using different scenarios, 
such as different grid sizes and different sensing ranges.
We also plan to run CG$k$CA under different incentive mechanisms and in real 
test beds.


\bibliographystyle{plain}
\bibliography{biblo}

\end{document}